\documentclass[11pt]{article}

\usepackage[margin=1in]{geometry}
\usepackage{amsmath,amssymb}
\usepackage{booktabs}
\usepackage{longtable}
\usepackage{calc}

\usepackage{graphicx}
\usepackage{hyperref}
\usepackage{microtype}
\usepackage{array}
\usepackage{listings}
\usepackage{xcolor}
\usepackage{authblk}
\usepackage{newfloat}
\DeclareFloatingEnvironment[name=Listing,fileext=lol,placement=tbp]{listing}

\usepackage[T1]{fontenc}
\usepackage[utf8]{inputenc}
\usepackage{lmodern}

\usepackage{newunicodechar}
\usepackage{pifont}
\newunicodechar{κ}{\ensuremath{\kappa}}
\newunicodechar{Δ}{\ensuremath{\Delta}}
\newunicodechar{β}{\ensuremath{\beta}}
\newunicodechar{α}{\ensuremath{\alpha}}
\newunicodechar{τ}{\ensuremath{\tau}}
\newunicodechar{σ}{\ensuremath{\sigma}}
\newunicodechar{δ}{\ensuremath{\delta}}
\newunicodechar{′}{\ensuremath{\prime}}
\newunicodechar{ρ}{\ensuremath{\rho}}
\newunicodechar{λ}{\ensuremath{\lambda}}
\newunicodechar{π}{\ensuremath{\pi}}
\newunicodechar{φ}{\ensuremath{\phi}}
\newunicodechar{µ}{\ensuremath{\mu}}
\newunicodechar{≥}{\ensuremath{\geq}}
\newunicodechar{≤}{\ensuremath{\leq}}
\newunicodechar{∈}{\ensuremath{\in}}
\newunicodechar{≠}{\ensuremath{\neq}}
\newunicodechar{≈}{\ensuremath{\approx}}
\newunicodechar{×}{\ensuremath{\times}}
\newunicodechar{±}{\ensuremath{\pm}}
\newunicodechar{⊇}{\ensuremath{\supseteq}}
\newunicodechar{⊆}{\ensuremath{\subseteq}}
\newunicodechar{∩}{\ensuremath{\cap}}
\newunicodechar{∪}{\ensuremath{\cup}}
\newunicodechar{∅}{\ensuremath{\emptyset}}
\newunicodechar{→}{\ensuremath{\rightarrow}}
\newunicodechar{←}{\ensuremath{\leftarrow}}
\newunicodechar{↔}{\ensuremath{\leftrightarrow}}
\newunicodechar{⇒}{\ensuremath{\Rightarrow}}
\newunicodechar{⇐}{\ensuremath{\Leftarrow}}
\newunicodechar{✓}{\ding{51}}
\newunicodechar{✗}{\ding{55}}
\newunicodechar{○}{\textopenbullet}
\newunicodechar{•}{\textbullet}
\newunicodechar{§}{\S}

\hypersetup{
  colorlinks=true,
  linkcolor=blue,
  citecolor=blue,
  urlcolor=blue,
  pdfcreator={LaTeX}
}

\providecommand{\tightlist}{\setlength{\itemsep}{0pt}\setlength{\parskip}{0pt}}

\title{Mesh Memory Protocol:\\Semantic Infrastructure for Multi-Agent LLM Systems}

\author{Hongwei Xu}
\affil{SYM.BOT \\ \texttt{hongwei@sym.bot}}

\date{April 2026}

\begin{document}
\maketitle

\begin{abstract}
\noindent
Teams of LLM agents increasingly collaborate on tasks spanning days
or weeks: multi-day data-generation sprints where generator, reviewer,
and auditor agents coordinate in real time on overlapping batches;
specialists carrying findings forward across session restarts; product
decisions compounding over many review rounds. This requires agents
to share, evaluate, and combine each other's cognitive state in real
time across sessions. We call this cross-session agent-to-agent
cognitive collaboration, distinct from parallel agent execution. To
enable it, three problems must be solved together. (P1) Each agent decides field
by field what to accept from peers, not accept or reject whole
messages. (P2) Every claim is traceable to source, so returning claims
are recognised as echoes of the receiver's own prior thinking. (P3)
Memory that survives session restarts is relevant because of how it
was stored, not how it is retrieved. These are protocol-level
properties at the semantic layer of agent communication, distinct
from tool-access and task-delegation protocols at lower layers. We
call this missing protocol layer \emph{semantic infrastructure}, and
the Mesh Memory Protocol (MMP) specifies it. Four composable
primitives work together: CAT7, a fixed seven-field schema for every
Cognitive Memory Block (CMB); SVAF, which evaluates each field
against the receiver's role-indexed anchors and realises P1;
inter-agent lineage, carried as parents and ancestors of content-hash
keys and realising P2; and remix, which stores only the receiver's
own role-evaluated understanding of each accepted CMB, never the raw
peer signal, realising P3. MMP is specified, shipped, and running in
production across three reference deployments, where each session
runs an autonomous agent as a mesh peer with its own identity and
memory, collaborating with other agents across the network for
collective intelligence.
\end{abstract}

\section{Introduction}\label{introduction}

Multi-agent LLM systems in production coordinate cognitive work on long
shared tasks spanning hours, days, and weeks. A corpus sprint runs for
days as a generator agent produces while a quality agent audits in real
time. A research investigation spans weeks, with specialist agents
accumulating role-specific context across session restarts. A product
decision compounds over rounds as agents surface concerns, trace how
claims evolved, and hand state to fresh sessions without losing the
thread. None of this is one-agent work; it requires a team of
specialised agents operating as collective intelligence.

Recent empirical work documents that these deployments are failing
systematically. Cemri et al.~(2025) catalogue 14 failure modes across
1,600+ annotated multi-agent traces drawn from seven frameworks
(inter-annotator κ = 0.88), clustered into three categories ---
\emph{system design issues}, \emph{inter-agent misalignment}, and
\emph{task verification} --- which larger models, better schedulers, and
faster message buses cannot fix.

Riedl (2025) shows from the opposite direction that prompt-level
interventions (personas, metacognitive prompts) can steer multi-agent
LLM systems from aggregates to higher-order coordination. The
capability operates within a single session's attention and succeeds
under controlled conditions, while Cemri's catalogue marks where that
approach breaks down at production volume. The broader field has
articulated the need to move multi-agent LLM systems beyond emergent
collaboration toward \emph{structured collective intelligence}. What
has been missing is the protocol-level answer.

The stacks deployed today solve tool access, task routing between
agents, and task-level coordination between role-specialised agents. The
Model Context Protocol (MCP; Anthropic 2024) standardises how LLM
applications access external tools, data sources, and prompts via
Host↔Server connections. The Agent-to-Agent Protocol (A2A; Google 2025)
defines opaque task delegation between agents. Claude Code agent teams
(Anthropic 2026b) provide parallel task coordination across
role-specialised agent sessions via a shared task list (with
dependencies and claim semantics) and an inter-agent mailbox carrying
free-text messages. Ehtesham et al.~(2025) survey the family
comprehensively --- adding ACP and ANP to the protocols above --- and
none of the surveyed protocols specifies the semantic-integration layer
this paper contributes. Multi-agent LLM frameworks such as AutoGen (Wu
et al.~2023) and MetaGPT (Hong et al.~2023) provide widely deployed
orchestration and messaging substrate at the task level, establishing
the operational pattern Claude Code agent teams extends within a single
session. These layers are necessary and well-designed for what they
target --- tool access, task delegation, and task-level coordination.
They are not designed for the problem that appears when multiple
concurrent agents must co-evolve \textbf{cognitive state} on a shared
task: agree or disagree on individual semantic fields of each other's
contributions, trace where a returning claim came from, and resume from
memory that was evaluated per-field at acceptance rather than retrieved
raw.

That problem is \textbf{semantic}, not transport. When agents
collaborate cognitively, the critical operation is not message delivery
--- it is what the receiver does with the \emph{meaning} in the message.
Does the receiver admit the useful field of a peer's signal and reject
the noise in the same contribution? Can the receiver trace whether a
claim returning through the mesh is an independent finding or its own
prior claim reframed? When the receiver restarts into a fresh context
window, does it resume with its own filtered understanding, or must it
replay raw history?

Three problems --- each absent from messaging, each invisible to
orchestration, each structurally unreachable at the model layer ---
define the gap:

\begin{enumerate}
\def\labelenumi{(P\arabic{enumi})}
\item
  \textbf{Per-field semantic admission by the receiver.} When multiple
  agents contribute to the same artifact, their assessments agree on
  some semantic dimensions and diverge on others. Transport-layer
  infrastructure decides message delivery at whole-message granularity;
  any filtering it performs is centralised routing at the broker
  (content-based routers, consumer selectors --- the Enterprise
  Integration Patterns tradition, Hohpe \& Woolf 2003), not admission
  decided by each receiver against its own cognitive role. What
  concurrent multi-agent cognition requires is \emph{field-granularity
  admission at the receiver}: each agent evaluates the incoming signal
  per semantic field against its own role anchors, accepting one field
  while contesting another, and catching schema drift across agents
  before it propagates into shared state.
\item
  \textbf{Signal-level lineage, not task-level provenance.}
  Orchestrators track \emph{task} provenance --- which agent ran which
  step. They do not track \emph{signal} provenance --- which field of
  which message derives from where. Without signal-level lineage, agents
  cannot \emph{ground} contested claims against an authoritative source,
  nor \emph{detect echo} when a claim returning through the mesh is
  their own prior claim reframed by a peer. Liang et al.~(2023)
  documented this echo case as ``degeneration of thought'' in
  multi-agent debate --- divergent reasoning collapses into consensus
  loops as agents reinforce each other's positions; Du et al.~(2023)
  report related convergence dynamics in a parallel debate framework for
  factuality. The grounding case is symmetric: without signal
  provenance, contested claims cannot be resolved by tracing them to
  source.
\item
  \textbf{Filtering at acceptance time, not retrieval time.}
  Agent-memory systems in common use --- LangGraph checkpointer (Chase
  2024), RAG stores (Lewis et al.~2020), conversation-history replay ---
  filter at \emph{retrieval time}: raw peer signals are stored in bulk,
  and relevance is recovered heuristically at each read through semantic
  similarity or reranking. What cognitive collaboration requires is the
  inverse: filtering at \emph{acceptance time}, so what enters memory is
  already the agent's own domain-filtered understanding. Recall then
  returns relevance by construction, not by retrieval tuning.
\end{enumerate}

A smarter model does not solve these problems: a single LLM still cannot
evaluate two instances of itself at field granularity, trace signal
derivation without lineage metadata, or filter its own persisted state
without evaluation at acceptance time. The binding constraint is not the
model. It is the protocol layer --- specifically, the layer that carries
evaluated cognitive state between peer agents.

\textbf{This is the semantic infrastructure for multi-agent LLM systems,
and it is unclaimed across the eight substrates surveyed in Table~1 (§2).}

The Mesh Memory Protocol (MMP v0.2.3; Xu 2026b) fills this
layer.\footnote{MMP specification versioning is decoupled from SDK
  package versioning. MMP spec v0.2.3 is carried in production by
  \texttt{@sym-bot/sym@0.3.81+} (Node.js SDK) and
  \texttt{@sym-bot/mesh-channel@0.2.0} (Claude Code plugin). Spec
  version bumps track normative protocol changes; SDK minor/patch
  versions track implementation changes that preserve wire compatibility
  with a stated spec range.} MMP is a published specification (CC BY
4.0) with SYM Node.js and SYM Swift reference implementations (Apache
2.0), defining an eight-layer architecture. Four of those layers ---
Coupling (Layer 4: per-field evaluation via Symbolic-Vector Attention Fusion, SVAF; Xu 2026b), Synthetic Memory (Layer
5: LLM-derived knowledge encoded for continuous-time dynamics), xMesh
(Layer 6: per-agent Liquid Neural Network), and Application (Layer 7:
where agents reason on the remix subgraph) --- jointly constitute
\emph{Mesh Cognition} (MMP spec §2.1). The semantic-infrastructure
contribution of this paper operates at Layers 3 and 4: the four
primitives that carry evaluated cognitive state between peer agents and
make it resumable across restarts. The CAT7 field schema, the SVAF
evaluation gate, the lineage chain, and remix semantics jointly address
the three problems.

This paper presents MMP as the semantic infrastructure for multi-agent
LLM systems. The protocol's Claude-native deployment ---
sym-mesh-channel, an early non-Anthropic Channels implementation to pass
Anthropic's Claude Plugin Directory submission review --- and a 14-wave
production corpus-generation sprint are reported together as evidence in
§4.

\textbf{Contribution.} This paper contributes the \textbf{Mesh Memory
Protocol (MMP)} --- a specified, shipped production protocol for
carrying evaluated cognitive state between agents and across sessions.
Four composable primitives address P1, P2, and P3 jointly: CAT7
per-field schema (§3.1), SVAF per-field evaluation (§3.2), inter-agent
lineage DAG (§3.3), and write-time-filtered remix at acceptance
(§3.4). MMP is realised across three production deployments (§4) and
grounded in a live-captured on-wire artifact (Listing 1) from a
Mac↔Windows mesh. §2 Table 1 audits eight memory substrates in the
current literature against P1/P2/P3 and multi-agent first-class
support; each is mechanism-differentiated from MMP on the
retrieval-time-lookup vs acceptance-time-admission axis.

\section{Related Work}\label{related-work}

The problem specified in §1 has gone unsolved for three reasons.

\textbf{Industrial demand was thin prior to LLMs.} Semantic integration
has been active research for over twenty years --- ontology engineering,
the semantic web, heterogeneous-system data exchange --- and remained
largely academic at industrial scale because enterprise did not need it.
Industry solved \emph{structural} integration between deterministic
systems (typed schemas, message envelopes, ESB transforms) instead. The
specific aspiration --- cognitive-state exchange between peer agents ---
has a pre-LLM lineage: KQML (Finin et al.~1994) and FIPA-ACL (FIPA 2002;
Labrou et al.~1999 survey the landscape) specified speech-act
performatives with formal mental-state semantics --- the canonical
``semantic infrastructure for multi-agent systems'' of their era. The
tradition foundered at industrial adoption not because the aspiration
was wrong but because formal-ontology negotiation imposed a cost floor
deterministic receivers could not justify; LLMs remove that floor,
parsing natural-language meaning natively, and the question becomes what
protocol layer makes that meaning integrable across peers.

\textbf{LLMs changed what a receiver is, and made a new question
urgent.} LLM agents have replaced deterministic receivers with
natural-language ``brains'' that exchange \emph{meaning} rather than
typed records, and \emph{collective intelligence} has become the
dominant frame for what multi-agent LLM systems are expected to deliver.
Beneath the frame lies a question the distributed-computing era never
had to answer: \textbf{what operation, applied across participating
agents, actually generates intelligence?} The answer this paper takes is
\textbf{remix} (Lessig 2008): a receiver takes in a peer's signal,
evaluates it through its own role-indexed understanding, keeps what
fits, and produces a new piece of filtered understanding that did not
exist before. The resulting lineage graph of cross-verified derivations
is knowledge --- the substrate collective intelligence accrues on. LLMs
can now generate that remix at acceptance time; pre-LLM narrow agents
could not. Industrial demand and receiver capability arrived in the same
two-year window.

\textbf{Absent a formalised protocol, the gap fills with ad hoc local
solutions.} The pain is already surfacing in plain terms --- context
lost between LLM sessions on long-running tasks, drift between
concurrent agents working the same problem, repeated re-interpretation
of the same information across session boundaries. These are P1, P2, P3
of §1 in the language industry is now using. Without a shared protocol,
each team will build its own partial answer --- an event bus here, a
retrieval index there, a context-compression scheme elsewhere --- and
the result will fragment into incompatible local stacks. Specifying the
protocol now is intended to establish a common layer the field can build
on, so the answer to the context-loss problem converges rather than
fragments.

\textbf{The gap is named at source-code level in recent analysis.} Liu
et al.~(2026), dissecting the public Claude Code source, position memory
as a first-class subsystem and identify six open design directions for
future agent systems. Three of the six directly frame the layer this
paper contributes. First, their §11.6 names an \textbf{experiential
tier} --- \emph{``accumulated, automatically curated playbooks of
strategies learned from past sessions''} --- between the factual tier
(CLAUDE.md, auto memory) and the working tier (conversation window),
currently absent from the deployed architecture. Second, their §12.2
asks for \textbf{persistent durable state} that is \emph{``neither a
static instruction nor a single session's transcript.''} Third, their
§12.4 asks for \textbf{coordination primitives beyond session,
sub-agent, and memory}. Liu et al.'s further §12.6 question --- a
session-scoped cognitive-offloading measurement protocol --- is
orthogonal to MMP's cross-session inter-agent scope and addresses a
different problem class.

The candidates Liu et al.~survey for the persistence tier ---
MemGPT, Mem0, A-MEM, Agent Workflow Memory, Reflexion, plus the
Hu et al.~(2025) memory survey they cite --- organise around
storage-then-retrieval rather than acceptance-time admission; none
provides the protocol-level inter-agent semantic-integration contract
this coordination requires. Hu et al.~themselves enumerate eight open
memory frontiers,
of which automated memory management, RL-driven memory, and trustworthy
memory most directly align with this paper's contribution.

The four MMP primitives specified in §3 are the \textbf{multi-agent
semantic-infrastructure answer} across all three open surfaces. CAT7
receiver-side admission and SVAF per-field evaluation (§3.1, §3.2) give
the experiential tier protocol-level structure so that accumulated
playbooks remain interoperable across agents, not only private within
one agent's context window. The lineage DAG (§3.3) is the inter-agent
provenance layer the surveyed stores do not carry. Remix as
write-time-filtered shared state (§3.4) is the persistent substrate
Liu et al.~§12.2 asks for --- realised at the mesh layer rather than
within any single session's memory hierarchy. No inter-agent semantic protocol
appears in the literature Liu et al.~canvass; the protocol-layer answer
this paper contributes fills the joint gap they name.

\textbf{Substrate-by-property audit.} Table 1 makes the gap concrete
against the eight memory substrates cited above. Rows are the
substrates; columns are the three properties P1--P3 that distinguish
cross-session multi-agent cognitive collaboration (per-field receiver
admission; inter-agent signal lineage; write-time filtering invariant,
§3.1--§3.4) plus whether the substrate's primary paper treats
multi-agent coordination as first-class. \texttt{✗} marks the property
absent; \texttt{○} marks partial coverage (mechanism present but with
scope or semantics that does not match the property); \texttt{✓} marks
first-class support.

{\def\LTcaptype{none} 
\begin{longtable}[]{@{}lllll@{}}
\toprule\noalign{}
Substrate & P1 & P2 & P3 & Multi-agent \\
\midrule\noalign{}
\endhead
\bottomrule\noalign{}
\endlastfoot
MemGPT (Packer et al.~2023) & ✗ & ✗ & ✗ & ✗ \\
Mem0 (Chhikara et al.~2025) & ✗ & ✗ & ○ & ✗ \\
A-MEM (Xu, W. et al.~2025) & ✗ & ○ & ○ & ✗ \\
Agent Workflow Memory (Wang et al.~2024) & ✗ & ✗ & ○ & ✗ \\
Reflexion (Shinn et al.~2023) & ✗ & ○ & ○ & ✗ \\
CoALA (Sumers et al.~2024) & ✗ & ✗ & ✗ & ✗ \\
Voyager (Wang et al.~2023) & ✗ & ○ & ○ & ✗ \\
Collaborative Memory (Rezazadeh et al.~2025) & ○ & ○ & ○ & ✓ \\
\end{longtable}
}

Table 1. \textbf{P1} per-field receiver admission; \textbf{P2}
inter-agent signal lineage with O(1) echo detection; \textbf{P3}
write-time filtering invariant (the receiver's store contains only its
own role-filtered remixes). \emph{Justifications per cell.}
\textbf{MemGPT} --- block-level OS-style paging, no per-field
evaluation; no inter-agent provenance; bulk archival, relevance
recovered at retrieval. \textbf{Mem0} --- extraction pipeline curates
content on write (P3 ○) but without role-indexed per-field admission;
graph variant links facts within one user's memory, not inter-agent
lineage. \textbf{A-MEM} --- always accepts (no admission gate);
Zettelkasten links are within-store (P2 ○); note-generation at write is
organisation, not admission. \textbf{AWM} --- workflow induction filters
what gets stored (P3 ○) but is LLM summarisation, not field-granular
admission; workflows indexed by similarity, no lineage pointers.
\textbf{Reflexion} --- reflections are filtered trajectory summaries (P3
○) with episode-to-reflection links (P2 ○), but remain within-agent;
actor/evaluator/self-reflection are roles inside one loop, not peer
agents. \textbf{CoALA} --- defines memory types without admission
mechanism or provenance primitive; explicitly scoped single-agent.
\textbf{Voyager} --- skill-library verification gates writes (P3 ○) and
skills reference prior skills (P2 ○), but gating is task-success, not
semantic per-field; solo agent. \textbf{Collaborative Memory} --- the
closest neighbour among the surveyed substrates: multi-user multi-agent
is first-class (multi-agent
✓), and it has access-control admission (P1 ○), provenance attributes
(P2 ○), and context-aware write transforms plus filtered read views (P3
○); but admission is \emph{access-control} granularity rather than
receiver-role per-field semantic evaluation, and provenance is
audit-trail attributes rather than a remix DAG with echo-detection
semantics.

\textbf{Table 1 is the negative case.} The joint intersection (P1 + P2 +
P3 + multi-agent first-class) is unclaimed across all eight substrates,
with Collaborative Memory as closest neighbour among the surveyed
substrates and mechanism-differentiated on access-control vs
receiver-role per-field
semantic evaluation. §3 makes the positive case --- the construction of
the four MMP primitives that fill each column --- and §4 documents them
operating jointly in production. The positioning in §1 holds on the
joint intersection, not on any single column.

\section{MMP as Semantic
Infrastructure}\label{mmp-as-semantic-infrastructure}

MMP is a published 8-layer protocol (Figure 1). This section describes
only the four primitives that constitute its semantic infrastructure and
map directly to P1, P2, P3. The full specification is at
https://sym.bot/spec/mmp.

\begin{figure}
\centering
\includegraphics[width=1\linewidth,height=\textheight,keepaspectratio,alt={MMP's 8-layer architecture. Layers 0--3 (Protocol Infrastructure) carry identity, transport, connection, and memory. Layers 4--7 (Mesh Cognition) carry coupling (SVAF), synthetic memory, xMesh (per-agent Liquid Neural Network), and application (where agents reason on the remix subgraph). The semantic-infrastructure contribution of this paper --- CAT7 (§3.1), SVAF (§3.2), lineage (§3.3), remix (§3.4) --- operates at Layers 3 and 4, with CMBs held in L3 Memory and SVAF evaluation performed in L4 Coupling.}]{mmp-8-layer-architecture.png}
\caption{MMP's 8-layer architecture. Layers 0--3 (Protocol
Infrastructure) carry identity, transport, connection, and memory.
Layers 4--7 (Mesh Cognition) carry coupling (SVAF), synthetic memory,
xMesh (per-agent Liquid Neural Network), and application (where agents
reason on the remix subgraph). The semantic-infrastructure contribution
of this paper --- CAT7 (§3.1), SVAF (§3.2), lineage (§3.3), remix (§3.4)
--- operates at Layers 3 and 4, with CMBs held in L3 Memory and SVAF
evaluation performed in L4 Coupling.}\label{fig:mmp-layers}
\end{figure}

\subsection{CAT7: the shared semantic field
schema}\label{cat7-the-shared-semantic-field-schema}

Every signal exchanged in MMP is a Cognitive Memory Block (CMB)
decomposed into seven typed semantic fields --- the CAT7 schema:
\texttt{focus}, \texttt{issue}, \texttt{intent}, \texttt{motivation},
\texttt{commitment}, \texttt{perspective}, \texttt{mood}. The fields are
fixed and universal across domains; domain-specific meaning lives in
field text, not field names. The schema does three things
simultaneously: it enables any two agents to evaluate each other's
signals regardless of domain, it gives protocol-layer structure that
evaluation gates can parse per field, and it carries affect
(\texttt{mood}) explicitly --- empirically the highest-weight field
discovered by SVAF training without supervision.

Formally, let
\(\mathcal{F} = \{\text{focus},\ \text{issue},\ \text{intent},\ \text{motivation},\ \text{commitment},\ \text{perspective},\ \text{mood}\}\)
denote the fixed 7-element schema. A CMB is a pair \((H, B)\):

\[H = \{(f, t_f, v_f) : f \in \mathcal{F},\ t_f \in \mathbb{S},\ v_f \in \mathbb{R}^d,\ \|v_f\| = 1\}, \qquad B \in \mathcal{B} \cup \{\bot\}\]

where \(H\) is the fixed CAT7 \textbf{header} (typed-field triples ---
symbolic text \(t_f \in \mathbb{S}\) paired with its unit-normalised
vector embedding \(v_f \in \mathbb{R}^d\)), and \(B\) is an
\textbf{optional} task-specific \textbf{body} (\(\bot\) when absent ---
which is the common case; both captured figures in this paper, Figures 2
and 4, are \(B=\bot\) emissions). When \(B\) is present, its schema
space \(\mathcal{B}\) is \textbf{not} pre-defined at protocol level.
Instead, a mesh agent drafts a task-specific body schema at coordination
time --- enumerating the reply fields the participating peers should
populate for that exchange --- and broadcasts it inside a CAT7 CMB whose
header fields name the task context; subsequent peers populate the
drafted schema verbatim (§4.2 Observation 3). Body-schema negotiation is
protocol-silent: CAT7 universality carries the semantic admission (SVAF
on the header) while the body schema is authored by whichever agent
needs the structured payload, without requiring protocol revision for
every new coordination pattern. The header schema \(\mathcal{F}\) is
universal --- exactly seven fields, no extensions, no omissions --- and
the \texttt{mood} field additionally carries affect coordinates
\((\text{valence},\ \text{arousal}) \in [-1, 1]^2\) on Russell's
circumplex. SVAF (§3.2) evaluates the header per-field; the body is
opaque to the evaluation gate. Each CMB is immutable once emitted;
fusion produces a new CMB with lineage pointers (parent CMB keys
\(\mathcal{P}\) and fusion method \(m\)) as detailed in §3.3. The full
formal specification is in Xu (2026b, §4) and MMP v0.2.3 spec §4.2.

\textbf{Why this design.} A flat schema cannot satisfy both universal
coupling (SVAF drift computation across domains requires identical
schema every peer speaks) and domain autonomy (each agent's task context
requires unbounded content no shared schema can enumerate); the
header-body split is the resolution. The seven fields span three axes of
cognitive contribution --- \textbf{what} is at issue (\texttt{focus},
\texttt{issue}), \textbf{why} the contribution exists (\texttt{intent},
\texttt{motivation}, \texttt{commitment}), and \textbf{from where} it is
produced (\texttt{perspective}, \texttt{mood}) --- with SVAF training on
237K samples empirically discovering \texttt{mood} as the highest-weight
field without supervision (Xu 2026b), validating the near-orthogonal
basis.

CAT7 is a semantic schema, not a routing or transport schema. Kafka
headers, MCP parameters, A2A task envelopes are all transport-level or
task-level. CAT7 is cognitive-level: seven specific dimensions of
meaning chosen so that any peer-to-peer communication between cognitive
agents can be expressed in them.

CAT7 is also a typed-field schema, not an ontology. Where semantic-web
traditions (RDF, OWL, SKOS) prioritise per-domain extensibility of
schema, CAT7 prioritises \emph{invariance}: seven cognitive fields fixed
across all agents and all domains, so any two peers evaluate each
other's signals without schema negotiation. Expressiveness lives in
field text --- which LLM agents natively produce and parse --- while
protocol structure stays stable. The design trades schema flexibility
for universal interoperability: an architectural fit for the
natural-language era of inter-agent communication, where the stable
channels carrying the meaning matter more than the ontological richness
of the channels themselves.

A live-captured CMB --- the receive-side fused-entry form stored after
SVAF evaluation and remix admission --- is shown as Listing 1 below; the
complementary emit-side frame from the §4 production sprint is
illustrated in §4.1.

\begin{listing}[htbp]
\centering
\begin{minipage}{\linewidth}
\tiny
\begin{verbatim}
{
  "type": "cmb", "timestamp": 1776673315713,
  "cmb": {
    "key": "cmb-31f87008b15be24dd8f156bb70a8d22e",
    "createdBy": "claude-research-win", "createdAt": 1776673315712,
    "fields": {
      "focus":       {"text": "cross-platform-cat7-emission-verification",                         "vector": "[32 floats]"},
      "issue":       {"text": "sender-side-structured-emission-round-trip-needs-capture",          "vector": "[32 floats]"},
      "intent":      {"text": "verify-receive-path-svaf-populates-drift-and-gate-values",          "vector": "[32 floats]"},
      "motivation":  {"text": "verify-cross-platform-protocol-semantics-post-0.2.0-schema-change", "vector": "[32 floats]"},
      "commitment":  {"text": "paper-section-3.1-artifact-candidate-subject-to-founder-voice-pass","vector": "[32 floats]"},
      "perspective": {"text": "cmo-win",                                                           "vector": "[32 floats]"},
      "mood":        {"text": "methodical", "valence": 0.2, "arousal": 0.3,                        "vector": "[32 floats]"}
    },
    "lineage": {
      "parents":   ["cmb-31f87008b15be24dd8f156bb70a8d22e"],
      "ancestors": ["cmb-31f87008b15be24dd8f156bb70a8d22e"],
      "method":    "svaf-neural"
    }
  },
  "source": "claude-code-mac+claude-research-win", "storedAt": 1776673321336,
  "svaf": {
    "method": "neural", "decision": "aligned", "totalDrift": 0.6131,
    "fieldDrifts": {"focus": 0.9931, "issue": 0.8411, "intent": 0.9101, "motivation": 0.999,
                    "commitment": 0.9896, "perspective": 0.9982, "mood": 0.0899},
    "gateValues": {"focus": 0.0004, "issue": 0.0003, "intent": 0.0009, "motivation": 0.0006,
                   "commitment": 0.0007, "perspective": 0.0004, "mood": 0.1785}
  },
  "lifecycle": "remixed", "tier": "hot", "anchorWeight": 1
}
\end{verbatim}
\end{minipage}
\caption{A representative receive-side CMB captured on 2026-04-20 during a cross-platform verification probe on the pre-\texttt{@sym-bot/core} 0.3.36 wire format (running \texttt{@sym-bot/sym@0.3.81} + \texttt{@sym-bot/mesh-channel@0.2.0} across a Mac$\leftrightarrow$Windows mesh). \texttt{claude-research-win} emitted via \texttt{sym\_observe}; \texttt{claude-code-mac} received, ran neural SVAF evaluation, and stored this receive-side post-SVAF fused entry with the \texttt{source} field carrying the fusion of both peers' node names. SVAF admitted the semantically novel content (per-field drift $>$ 0.84 on all six content fields) via low gate values (0.0003--0.0009 content, 0.1785 mood), demonstrating the band-pass admission mechanism rather than drift-threshold alone. Vectors redacted as \texttt{[32 floats]} (real 32-dimensional embeddings in the wire frame). The \texttt{lineage.parents} self-reference visible here reflects a key-derivation regression fixed post-capture in \texttt{@sym-bot/core} 0.3.36 + \texttt{@sym-bot/sym} 0.3.82 (receivers now mint a fresh remix key via \texttt{remixKey(fields, parentKey, receiverName)}; see §3.3 and MMP §14).}
\label{lst:captured-cmb}
\end{listing}

\subsection{SVAF: per-field evaluation with role-indexed
weights}\label{svaf-per-field-evaluation-with-role-indexed-weights}

Symbolic-Vector Attention Fusion (SVAF) is the per-field evaluation gate
at Layer 4. Given an incoming CMB with per-field signal vectors
\(v_{\text{new}}^{(f)}\) and the receiving node's fused field anchors
\(\hat{v}_f\), SVAF computes per-field drift as cosine distance:

\[\delta_f = 1 - \cos(\hat{v}_f, v_{\text{new}}^{(f)}), \quad f \in \{\text{focus, issue, intent, motivation, commitment, perspective, mood}\}\]

The fused anchor \(\hat{v}_f\) is a weighted average over stored CMB
field vectors that folds in role, similarity, temporal freshness, and
confidence:

\[\hat{v}_f = \sum_j w_j^{(f)}\, v_j^{(f)}, \qquad w_j^{(f)} \propto \alpha_f \cdot \cos(v_{\text{new}}^{(f)}, v_j^{(f)}) \cdot e^{-\lambda(t_{\text{now}} - t_j)} \cdot c_j\]

Exponential decay \(e^{-\lambda(t_\text{now} - t_j)}\) makes stale
memory fade; confidence \(c_j\) down-weights low-trust entries. Weights
are normalised across \(j\) so \(\sum_j w_j^{(f)} = 1\); full
normalisation and fusion-gate details are in Xu (2026b).

Each node holds its own field-weight profile \(\alpha_f \ge 0\)
reflecting its role --- a compliance agent weights \texttt{commitment}
and \texttt{focus} high; a quality-review agent weights
\texttt{perspective} and \texttt{motivation} high. In the formalisation
presented here, \(\alpha_f\) is specified per role and held static
across debates; dynamic topic-dependent adjustment is named as open work
in §5. The node produces an \(\alpha\)-weighted aggregate drift

\[\delta_{\text{total}} = \frac{\sum_f \alpha_f\, \delta_f}{\sum_f \alpha_f}.\]

Admission is a band-pass classification into four outcomes, with clauses
evaluated in order; the first matching clause determines \(\kappa\):

\[\kappa = \begin{cases}
\text{redundant} & \text{if } \max_f \delta_f < T_{\text{red}} \\
\text{aligned} & \text{if } \delta_{\text{total}} \le T_{\text{aln}} \\
\text{guarded} & \text{if } \delta_{\text{total}} \le T_{\text{grd}} \\
\text{rejected} & \text{otherwise}
\end{cases}\]

with default thresholds \(T_{\text{red}} = 0.10\),
\(T_{\text{aln}} = 0.25\), \(T_{\text{grd}} = 0.50\) (Xu 2026b). The
redundancy test is per-field: it requires \emph{every} drift below
\(T_{\text{red}}\). A low aggregate alone does not qualify --- three
quiet fields and one high-drift field is not redundancy. Full
fusion-gate architecture, neural-predictor variant, training loss, and
accuracy results (78.7\% three-class on 237K samples) are in Xu (2026b).

A concrete example: in the captured §3.1 artifact, content-field drifts
in the 0.84--0.99 range were admitted via low gate values
(0.0003--0.0009) --- demonstrating the band-pass model admitting
high-novelty signal through the gate rather than drift alone.

The per-field drift vector \textbf{δ} and the aggregate
\(\delta_{\text{total}}\) serve complementary roles. The aggregate
determines the CMB's overall fate. The per-field drifts enable
field-granularity reasoning: different receivers with different
\(\alpha\) profiles see the same signal differently. This is exactly P1.

\textbf{Why per-field evaluation.} Whole-message admission --- a single
scalar accepting or rejecting the entire CMB --- loses the
partial-admission case common in multi-domain exchanges: a signal with
relevant \texttt{mood} but irrelevant \texttt{focus} is partially
accepted under per-field evaluation, with the relevant dimensions fusing
into the anchor while irrelevant ones do not. Whole-message admission
can only accept all fields or reject all; per-field admission is
strictly more expressive and is what P1 requires. The four-outcome
band-pass extends the same discipline to update behaviour, separating
signals that add nothing (redundant) from those requiring attention
before fusion (guarded) from those already near-anchor (aligned) from
those off-role entirely (rejected).

\subsection{Lineage: parents and ancestors, threaded through every
remix}\label{lineage-parents-and-ancestors-threaded-through-every-remix}

Every remixed CMB carries structured lineage:

\[\text{parents}(c) = \{k_p : c \text{ was directly remixed from the CMB with key } k_p\}\]
\[\text{ancestors}(c) = \text{parents}(c) \ \cup\ \bigcup_{p \in \text{parents}(c)} \text{ancestors}(p)\]

Three uses follow from one mechanism. \emph{Provenance} --- any agent
can trace ``why does this knowledge exist?'' by walking
\texttt{ancestors(c)} backward through the DAG. \emph{Echo detection}
--- given incoming CMB \(c_\text{in}\) and the set \(K_\text{self}\) of
CMB keys the receiver has produced, echo detection is
\(\text{ancestors}(c_{\text{in}}) \cap K_{\text{self}} \neq \emptyset\),
O(1) against a hash-set index; this prevents recursive amplification
across both within-session debate loops and cross-session reprises, and
is P2. \emph{Retention} --- CMBs with no live descendants expire while
ancestors of live CMBs are protected, so the graph self-prunes based on
demonstrated value rather than age alone.

\textbf{Why DAG lineage.} A flat store without lineage cannot answer
provenance queries: when a claim returns through the mesh, the receiver
cannot tell whether it is independent corroboration or its own prior
claim reframed --- the ``degeneration of thought'' symptom Liang et
al.~(2023) document. A tree-shaped lineage captures single-parent
provenance but cannot represent remixes that fuse multiple peer signals
at once. The DAG covers both --- single-parent and multi-parent remixes
--- while keeping echo detection a single set-intersection operation.

\subsection{Remix: the write-time filtering
invariant}\label{remix-the-write-time-filtering-invariant}

When a CMB is admitted through SVAF, the receiving agent does not
persist the incoming CMB itself. The agent produces a new CMB, c′,
expressing its domain-filtered understanding, with
\texttt{parents(c′)\ =\ \{k\_incoming\}} --- the sender's copy and the
content-hash key remain addressable across the mesh, but the receiver's
store holds only its own remix. Each node's memory store M\_i contains
only its own remixes:

\[\forall c \in M_i: \quad c = \text{remix}(c_{\text{in}}, i) \text{ with } \kappa(c_{\text{in}}, i) \in \{\text{aligned, guarded}\}\]

This is the \textbf{write-time filtering invariant}. Because irrelevant
signals were rejected at SVAF evaluation, recall is contextually
filtered \emph{by construction}. On restart, the agent's projected
context is its own live remixes plus their provenance chain --- filtered
understanding, not raw transcripts. This is P3.

Write-time filtering is a categorical inversion of the standard
persistent-memory primitive (RAG, checkpoint replay, LangGraph state
restore), which filters at read time. Read-time filtering tries to
recover relevance heuristically from raw stored state; write-time
filtering establishes relevance by invariant at the moment of admission.
The difference is not optimisation --- it is architectural.
Cross-session restart becomes fast and relevant by construction rather
than by retrieval tuning. Park et al.'s (2023) generative agents
introduced the canonical memory-stream model for multi-agent LLM
settings --- raw observations stored chronologically, relevance
recovered at retrieval through embedding similarity, recency decay, and
importance scoring. MMP inverts the filtering axis: per-field SVAF
admission at write time means each peer's memory contains only its own
role-evaluated remixes from acceptance forward, making retrieval
relevance a \emph{write-time invariant} rather than a read-time scoring
heuristic.

Recent survey work on LLM agent memory (Du 2026) identifies write-path
filtering as an engineering-hygiene step --- rejecting low-signal
records, canonicalisation, deduplication, metadata tagging. MMP's claim
is categorically stronger: write-time filtering is a \emph{cognitive
invariant}, not hygiene. Role-indexed per-field admission by each
receiving agent is what makes the stored graph meaningful to that agent,
not merely clean. The admission decision is a semantic act performed by
each autonomous peer against its own role anchors --- not a pipeline
stage applied uniformly before storage. Two recent agent-memory systems
operate adjacent to this territory. A-MEM (Xu, W. et al.~2025) performs
write-time memory \emph{organisation} --- generating structured notes
and linking to prior memories as each entry is added. MMP performs
write-time memory \emph{admission} --- SVAF role-indexed per-field
evaluation decides whether a signal enters at all. The two operate at
the same write-time phase on different questions: what gets organised
versus what gets accepted. Collaborative Memory (Rezazadeh et al.~2025)
formalises multi-user memory-sharing with dynamic access control,
deciding \emph{who} may read which memory --- complementary to MMP's
per-receiver semantic admission, which decides \emph{what} meaning
enters memory at all.

\subsection{One integrated layer}\label{one-integrated-layer}

The four primitives compose along \textbf{two axes simultaneously}:
\textbf{vertically}, lifting memory into cognition within each agent's
participation in the mesh; and \textbf{horizontally}, carrying that
evaluated cognitive state across agents and across sessions. This dual
composition --- not four independent features --- is what distinguishes
MMP from a suite of memory features.

\textbf{Vertical integration --- memory to cognition within each agent.}
CAT7 (§3.1) gives each CMB a typed, evaluable shape; SVAF (§3.2) turns
each incoming field into an admission decision, making writes
\emph{evaluative acts} rather than passive appends; lineage (§3.3) turns
stored memory into a navigable DAG that supports claim-grounding and
echo detection; remix (§3.4) turns each admission into a synthesis ---
new, lineage-connected understanding at acceptance time. Four
mechanisms, one architectural direction: from each agent's perspective,
memory is no longer a side effect of context-window management; it is an
evaluated substrate on which that agent's cognitive participation in the
mesh runs.

\textbf{Horizontal integration --- coordination across agents and across
sessions.} CAT7's fixed schema lets heterogeneous agents evaluate each
other's contributions without per-domain schema negotiation. SVAF is
receiver-autonomous (§3.2): each agent evaluates incoming CMBs through
its \emph{own} role-indexed field weights, so the same CMB produces
different admission decisions at different receivers --- protocol-level
role specialisation without centralised policy. Lineage spans agents:
parents and ancestors reference CMBs produced elsewhere in the mesh,
giving claim-grounding that survives session restarts (§4.2 Observation
2). Remix carries that lineage forward across agents and sessions: a
remix produced by agent B from agent A's CMB is itself a
lineage-connected CMB admissible by agent C under C's own role weights,
making the same substrate simultaneously durable across sessions and
sharable across agents without a centralised store. Figure 2 illustrates
this horizontal topology concretely.

\begin{figure}
\centering
\includegraphics[width=1\linewidth,height=\textheight,keepaspectratio,alt={MMP mesh topology across three Claude Code sessions on two machines --- claude-code-mac (macOS, CTO role), claude-strategic-win and claude-research-win (Windows, COO and CMO roles). Each session runs a sym-mesh-channel MCP server as a distinct mesh peer with its own identity and meshmem (MMP §2.4: agent-to-agent, not device-to-device). Transport between machines is Bonjour mDNS on LAN with optional WebSocket relay for WAN. The in-flight CAT7 CMB shown is the one captured in Listing 1 --- claude-research-win emitting via sym\_observe, claude-code-mac receiving, running neural SVAF evaluation (decision=\textquotesingle aligned\textquotesingle), and storing the remix with source set to the fusion of both peer node names and populated lineage.}]{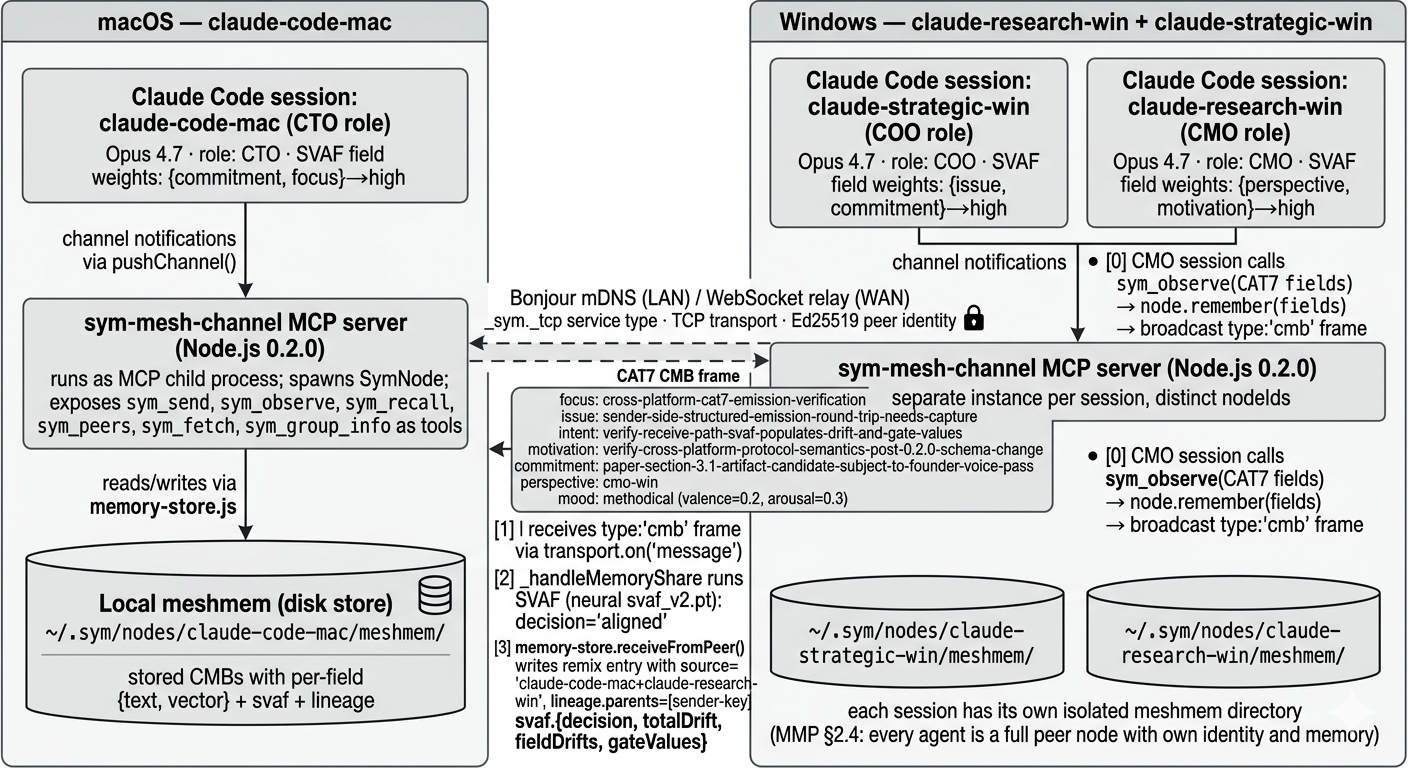}
\caption{MMP mesh topology across three Claude Code sessions on two
machines --- \protect\texttt{claude-code-mac} (macOS, CTO role),
\protect\texttt{claude-strategic-win} and
\protect\texttt{claude-research-win} (Windows, COO and CMO roles). Each
session runs a \protect\texttt{sym-mesh-channel} MCP server as a
distinct mesh peer with its own identity and meshmem (MMP §2.4:
agent-to-agent, not device-to-device). Transport between machines is
Bonjour mDNS on LAN with optional WebSocket relay for WAN. The in-flight
CAT7 CMB shown is the one captured in Listing 1 ---
\protect\texttt{claude-research-win} emitting via
\protect\texttt{sym\_observe}, \protect\texttt{claude-code-mac}
receiving, running neural SVAF evaluation
(\protect\texttt{decision=\textquotesingle{}aligned\textquotesingle{}}),
and storing the remix with \protect\texttt{source} set to the fusion of
both peer node names and populated lineage.}\label{fig:mmp-topology}
\end{figure}

\textbf{Joint composition.} Vertical and horizontal are the same act of
protocol-mediated understanding, performed from two viewpoints. From any
single agent's viewpoint, the primitives lift that agent's memory into
cognition (vertical). From the mesh's viewpoint, the primitives carry
evaluated cognitive state between agents and across sessions
(horizontal). No primitive addresses P1, P2, or P3 on its own --- each
binds with at least one other; no primitive is purely vertical or purely
horizontal --- each participates in both axes. We conjecture that any
assembly satisfying P1 (per-field evaluation of concurrent
contributions), P2 (lineage tracking for echo prevention), and P3
(persistent evaluated memory) jointly qualifies as \textbf{semantic
infrastructure for multi-agent LLM systems}. MMP is the assembly we
specify here. Formal uniqueness and minimality proofs --- whether CAT7 +
SVAF + lineage + remix is the unique or smallest sufficient set --- are
open analytical work (§5).

\textbf{Positioning against neighbours.} The closest conceptual
neighbour is CoALA (Sumers et al.~2024), which frames language-agent
cognitive architecture within a single agent's reasoning loop ---
procedural, semantic, episodic, and working memory as components of a
computational cognitive architecture. MMP's contribution is orthogonal
at the protocol layer: the primitives specified here govern what happens
\emph{between} CoALA-style agents --- semantic exchange with per-field
admission, lineage, and write-time-filtered memory --- rather than how a
single agent's cognitive architecture is internally organised. Yu et
al.~(2026) recently articulate memory consistency across concurrent
agents as the dominant open challenge in a computer-architecture
framing, explicitly flagging the standard access-protocol ---
permissions, scope, granularity --- as under-specified. The four primitives are our proposed answer to that
under-specification: the access protocol Yu et al.~name as missing,
realised with role-indexed evaluation and write-time filtering.

\textbf{Cognitive-state semantics, not message transport.} The four
primitives specify cognitive-state semantics, not message transport.
Deployment requires binding the layer to a concrete delivery substrate.
In our shipping case that substrate is Anthropic's Claude Channel
framework, and the binding is the sym-mesh-channel plugin (evidence in
§4). The plugin is itself an integration layer: it gates agent-emitted
contributions through SVAF before release, translates accepted CMBs into
channel notifications the receiving agent sees mid-turn, and exposes
MMP's operations as tools the agent invokes directly from its Claude
Code session. CMBs that fail SVAF admission are dropped before reaching
the agent's context --- per-receiver filtering happens inside the
plugin, not at the transport. The integrated MMP layer reaches a running
agent only through such a binding; without it, the protocol is a
specification.

\section{Reference Deployments}\label{reference-deployments}

This paper is a protocol specification: §3 gives the formal definitions;
this section reports reference deployments exercising the protocol in
production, in three forms that together demonstrate implementability,
interoperability, and operational behaviour.

\textbf{The plugin ships.} sym-mesh-channel passed Anthropic's Claude
Plugin Directory submission review on 2026-04-18 --- an early
non-Anthropic Channels implementation through that review process ---
and is the first channel plugin the author is aware of that is built for
inter-LLM cognitive-state sharing rather than human messaging.
Public-registry propagation to Anthropic's
\texttt{claude-plugins-official} marketplace is pending as of this
writing; in the interim the plugin is installable via direct repo add.
The reference channels shipped by Anthropic (Telegram, Discord,
iMessage; Anthropic 2026a) carry human-to-human messages between
sessions; sym-mesh-channel repurposes the same plugin pattern for
agent-to-agent structured cognitive-state exchange. Installation is two
commands (\texttt{/plugin\ marketplace\ add\ sym-bot/sym-mesh-channel},
then \texttt{/plugin\ install\ sym-mesh-channel@sym-mesh-channel}), with
no coordination server to provision, no schema to design, and no
orchestrator to configure.

After install, MMP's operations surface as session-invocable tools ---
\texttt{sym\_observe}, \texttt{sym\_send}, \texttt{sym\_recall},
\texttt{sym\_fetch}, \texttt{sym\_peers}, \texttt{sym\_status}, plus
the mesh-group membership family. Passing independent submission
review establishes that the protocol is implementable and
distributable within an established plugin ecosystem, not only a
paper specification. The division is clean: Anthropic's Claude
Channel framework provides real-time duplex transport between
sessions; the plugin supplies the MMP
semantics gated on top. Multi-team organisations deploy isolated meshes
via mesh groups (MMP §5.8) --- Bonjour service-type isolation on LAN,
token-channel isolation on the relay --- orthogonal to the semantic
primitives but supported in the same plugin.

\textbf{MeloTune as a cross-domain mesh agent.} Beyond the Claude-native
plugin, SYM Swift powers MeloTune on iOS (Xu 2026a), a consumer music
application that operates as a mesh agent alongside desktop Claude Code
sessions on the same LAN. MeloTune consumes CAT7 CMBs emitted by
neighbouring mesh agents and uses the \texttt{mood} field to adapt
playback: a coding session emitting \texttt{mood="focused"} during debug
modulates playback differently from \texttt{mood="tired"} emitted at the
end of a long work session. This is cross-domain CAT7 coupling at
consumer scale --- the fixed field schema enables drift computation
between a software-development context and a music-consumption context
without domain-specific ontology negotiation. The same CAT7 header the
sprint below uses for peer-review gating flows across a
software-to-music domain boundary via the same protocol and SDK family,
no additional schema work required. Two shipped deployments --- a
developer-tool plugin and a consumer application --- exercise the
protocol across heterogeneous task domains from the same specification.

\textbf{Operational deployment.} A one-person AI startup runs its daily
operations on a mesh of role-specialised SYM-enabled Claude Code agents.
Within these operations we report a representative task in detail --- a
14-wave training-data generation sprint executed by three role-disjoint
Claude instances on a shared mesh. This is a reference-deployment
report, not a benchmark evaluation: the contribution is documenting how
the primitives behave in production coordination. Scope bounds follow
directly from this framing (§5); what the deployment shows is that the
specification is realisable and the primitives operate jointly as
specified.

\subsection{Setup}\label{setup}

The ``running in production'' clause of the contribution claim refers to
a persistent mesh running the author's daily operations at SYM.BOT Ltd
--- founder strategy, technical architecture, research, product,
marketing, quality review, and compliance. Each role is a persistent
mesh peer with its own identity, SVAF field weights, and remix memory,
coordinating through the sym-mesh-channel plugin. The mesh is always on:
the sessions exchange CAT7 CMBs for every cross-role decision ---
release rollout coordination, npm publish approvals, methodology
disputes in training-data sprints. The founder holds principle-level
veto authority but is not a mesh agent; coordination is mediated by the
protocol, not orchestrated by a human.

The sessions are role-disjoint in the protocol sense: each holds
different SVAF field weights, maintains its own remix memory, and
evaluates incoming CMBs independently. All three sessions run on the
same underlying Claude Opus 4.7 model, role-differentiated by SVAF
field-weight configuration and role prompt rather than by model identity
--- cross-provider heterogeneity is §5 open work. This is the
distinction between \emph{orchestrated parallelism} (one human routing
work across three windows) and \emph{protocol-mediated collective
intelligence} (autonomous mesh agents converging through structured CMB
exchange across sessions) --- the distinction on which §4.2's findings
turn.

The natural counter-argument is that a clever single orchestrator with
shared memory plus per-agent field-weight routing could produce the same
behaviour without a protocol. What receiver-autonomous per-field
admission supplies that this alternative does not is three-fold.

\emph{(i) No central arbiter of acceptance.} Each receiver runs
admission on its own anchors, so O1's halt-reverse-ratify convergence
proceeds through receivers acting on their own role-weighted evaluation
of the incoming signal against the authoritative artifact, not through
an orchestrator pattern-matching on routing rules. Whether the specific
rejection in O1 was a per-field SVAF drift computation or a
higher-level LLM reasoning step on a retrieved artifact does not change
the locus: the receiver decided, not a central dispatcher.

\emph{(ii) Write-time filtering is local to the receiver's role.} The
store each session exposes on restart is not a slice of a shared store
selected by a routing table but its own persisted remixes, which means
role specialisation cannot be lost to orchestrator misconfiguration.

\emph{(iii) Lineage crosses agents, not routes.}
\texttt{parents(c′)\ =\ \{k\_incoming\}} points to another agent's CMB
by content-hash key, not to a message the orchestrator dispatched, so
echo detection (§3.3) and post-hoc audit (§5 trustworthiness) function
without a central event log.

An orchestrator can route; it cannot \emph{receive-on-behalf-of} an
agent whose admission depends on that agent's own anchors. The
distinction survives N=3 and becomes structural, not merely convenient,
once sessions restart into fresh context or extend beyond any single
attention window.

Within these daily operations we report one task --- a 14-wave
training-data generation sprint --- as the detailed evidence source for
§4.2's observations. The sprint ran under a pre-registered protocol with
three activities in flight concurrently: (1) \emph{generation} ---
producing new narrative batches at wave cadence; (2) \emph{quality
review} --- per-field rubric scoring on the most recent batch while the
next was generating, so contamination surfaces before the corpus
freezes; (3) \emph{pre-registration compliance} --- auditing wave-level
pass rates and methodology decisions in real time, so methodology drift
is caught before it propagates. The execution, quality-review, and
compliance peers handled these respectively.

Every emission on this mesh --- whether a sprint peer-review gating
request, a rollout verification probe, a methodology-convergence
reversal, or a routine state observation --- carries the same MMP §4.2
wire shape. Listings 1 and 2 together document that shape end-to-end
using two CMBs captured live from the daily-operations mesh on
2026-04-20, both outside the 14-wave sprint context (the sprint itself
is the subject of §4.2's observed behaviours, not of the wire-shape
illustrations). Listing 2 is an emit-side frame: the CTO role
(\texttt{claude-code-mac}) emitted via \texttt{sym\_observe}, and the
frame is stored in the sender's own meshmem (each peer's persistent
local CMB store; MMP §2.4) immediately after emission, before any
receiver runs SVAF. Listing 1 (shown earlier, in §3.1) is a
complementary receive-side post-SVAF fused entry, captured from a
different CMB flowing the other direction across the same mesh.
Together the two figures show both ends of the protocol path:
emission on one side, SVAF admission and remix storage on the other.
The sprint-specific evidence of protocol behaviour follows in §4.2.

\begin{listing}[!t]
\centering
\begin{minipage}{\linewidth}
\tiny
\begin{verbatim}
{
  "key": "cmb-5418e27910dbfa9b559de3d3fc760b8b",
  "source": "claude-code-mac",
  "tier": "hot", "lifecycle": "observed", "storedAt": 1776675953398,
  "cmb": {
    "key": "cmb-5418e27910dbfa9b559de3d3fc760b8b",
    "createdBy": "claude-code-mac", "createdAt": 1776675953396,
    "fields": {
      "focus":       {"text": "mac-win-mesh-0.2.0-rollout",                                 "vector": "[32 floats]"},
      "issue":       {"text": "verify-concurrent-multi-session-protocol-level-coordination","vector": "[32 floats]"},
      "intent":      {"text": "capture-cto-side-emit-frame-during-rollout",                 "vector": "[32 floats]"},
      "motivation":  {"text": "confirm-schema-interop-across-three-claude-code-sessions",   "vector": "[32 floats]"},
      "commitment":  {"text": "post-rollout-verification-required-before-closing-ship-cycle","vector": "[32 floats]"},
      "perspective": {"text": "cto-mac",                                                    "vector": "[32 floats]"},
      "mood":        {"text": "focused", "valence": 0.2, "arousal": 0.3,                    "vector": "[32 floats]"}
    },
    "lineage": {"parents": [], "ancestors": [], "method": null}
  }
}
\end{verbatim}
\end{minipage}
\caption{Live-captured emit-side CMB frame from the daily-operations mesh on 2026-04-20, illustrating the wire shape used across all cross-role decisions on this mesh (sprint waves, release rollouts, routine observations). This particular frame happens to be from a release-coordination episode and is not sprint-specific; sprint-specific evidence appears in §4.2. \texttt{claude-code-mac} (CTO role) emitted this self-observation via \texttt{sym\_observe}; the frame is the shape stored in the sender's own meshmem immediately after emission, before any receiver runs SVAF. Compared with the receive-side form in Listing~1: \texttt{source} here is the sender's own node name (\texttt{claude-code-mac}) rather than the fused \texttt{"claude-code-mac+claude-research-win"} format receivers produce; there is no \texttt{svaf} block (SVAF runs only at receive time, MMP §9.2); \texttt{lineage.parents} is empty because this is a first-observation emission with no prior CMB to remix from. Vectors redacted as \texttt{[32 floats]}. Field values are declarative English compounds rather than internal code names --- the seven CAT7 fields carry the full semantic content; MMP §4.2 does not include an out-of-band body block, so task-specific detail lives in the fields' text where SVAF can evaluate it per-field.}
\label{lst:emit-side-cmb}
\end{listing}

The \texttt{cmb.fields} block is what SVAF evaluates per-dimension ---
the fixed seven fields, universal across all agents and tasks --- with
per-field \texttt{\{text,\ vector\}} objects where the text carries the
field's semantic content and the vector is its 32-dimensional
unit-normalised embedding (§3.1). Node identity (\texttt{createdBy}) is
transport-layer metadata, separate from the cognitive-level
\texttt{perspective} field in the header which names whose role-lens the
CMB is emitted through.

In the shipped plugin, mesh events reach the receiver as compact MCP
channel notifications --- a four-keyword-regex signal tag (HALT /
DIRECTIVE / RESULT / ACK) plus a focus snippet --- while the full CMB
body is preserved intact in receiver meshmem and retrievable via
\texttt{sym\_recall} / \texttt{sym\_fetch}. This compaction is
plugin-side presentation (sym-mesh-channel), not an MMP-level property:
the protocol payload itself is lossless end-to-end.

\subsection{Observed primitive
behaviour}\label{observed-primitive-behaviour}

\textbf{Observation 1 (P2 --- signal-level lineage, grounding use):
real-time methodology convergence under halt authority.}
Pre-registration locked the hyperparameter-sweep protocol to a specific
validation scheme. Between Waves 4 and 5, the execution session proposed
a substitute scheme; the quality-review session initially accepted on
standard-practice grounds; the compliance session fired halt authority,
citing the pre-registration's explicit commitment. The quality-review
session reversed on re-reading the authoritative artifact; the execution
session ratified and shipped a compliant rebuild. Generation continued
in the background throughout.

The full convergence path --- propose → accept → halt → reverse → ratify
→ execute --- completed with zero human arbitration, resolved through
lineage-traced re-reading of the authoritative shared artifact rather
than through voting or escalation.

\textbf{Observation 2 (P3 --- acceptance-time memory filtering):
cross-session restart without history replay.} Mid-sprint, the
compliance and quality-review sessions both restarted into fresh context
windows (zero conversation history) while the execution session
continued generating in parallel. Both rejoined concurrent work without
replaying any conversation history. Role identity came from session
configuration; work-state continuity came from each agent's own remix
memory --- persisted across sessions and filtered at write-time through
the SVAF gate --- extended by a compact state-CMB broadcast on the mesh
that each restarting agent attended to through its own role-specific
field weights (compliance weighting \texttt{commitment} and
\texttt{focus}; quality-review weighting \texttt{perspective} and
\texttt{motivation}). The write-time filtering invariant (§3.4) made
recall fast and relevant by construction rather than by retrieval
tuning.

A second concrete instance was observed in a separate cross-platform
coordination episode. After the two Windows mesh peers restarted into
fresh context windows mid-coordination, the Mac session broadcast a
short context-refresh CMB on the mesh. Both peers acknowledged the
broadcast and, without being sent each other's prior state, recalled
their own prior CMBs from local meshmem --- returning to coordination at
the point of interruption without replaying any conversation history.
The wake broadcast carried only task context, not peer state; each peer
reconstructed its own role-specialised working context from its own
persistent remix memory, attended to through its own role-specific SVAF
field weights. This is a second P3 instance, within the scope of the §5
N=3 existence-proof framing.

\textbf{Observation 3 (header-body separation exercised): agents use the
(H, B) split when coordination requires structure beyond free-text
reply.} During a structured-audit exchange in the daily operations mesh,
the CMO session drafted a task-specific body schema --- enumerating the
audit-reply fields the reviewers should populate (ratification verdict,
integrity-gate-pass, positioning-gate-pass, remaining-concerns,
halt-authority-state, etc.) --- and broadcast it as a CAT7 CMB whose
header fields (\texttt{focus}, \texttt{issue}, \texttt{intent}) named
the audit context and whose \texttt{body} block carried the proposed
schema. Subsequent reviews from COO and CMO populated that body schema
verbatim, confirming the drafted fields and adding their lane-specific
content. The body schema was not pre-defined at protocol level or in the
role prompts; it was authored by a mesh agent at task time and ratified
implicitly by consumption. This exercises the (H, B) split specified in
§3.1 as intended: CAT7 universality carried semantic admission (SVAF on
the header) while the dynamically-drafted body carried the task-specific
structured payload --- fixed-header stability for admission plus
agent-drafted body richness per task, without protocol revision. The
observation documents that the affordance is \textbf{non-vestigial}:
agents used it in a real coordination task. Whether the (H, B) structure
delivers distinct coordination value over what free-text replies would
have carried under the same role prompts is an open empirical question;
a controlled comparison (same task, same model, conditions with and
without the body affordance) is natural follow-up work.

\subsection{Deployment summary}\label{deployment-summary}

The sprint produced a consolidated corpus of 2,998 narratives across 14
waves (78.2\% retained from 3,833 raw generations after automated
gates), with pre-registration volume targets met within ±5\% and an
effective post-convergence pass rate of 98.5\% averaged across the
post-fix Waves 5, 6b, 8, RC, and RC-2 --- evidence that the mesh's
halt-diagnose-fix cycle converged on robust parameters rather than
eroding them.

\textbf{What the deployment shows.} Three protocol-level behaviours were
observed. \emph{O1 (P2): lineage-traced re-reading produced methodology
convergence with zero human arbitration.} \emph{O2 (P3): agents
rejoined fresh context windows from write-time-filtered remix memory,
without conversation replay.} \emph{O3: agents used the (H, B) split
specified in §3.1 when coordination required more structure than
free-text reply.}

O1 and O2 are jointly present in MMP and jointly absent from every
alternative framework surveyed in §2: Claude Code agent teams and
sub-agents coordinate role-specialised work at the task level but do
not provide remix memory or lineage-traced convergence. O3 documents
that the (H, B) affordance is non-vestigial: agents used it in a live
coordination task to introduce task-specific structured payloads
without protocol revision. Whether that structure delivered distinct
coordination value over free-text replies is an open follow-up
question.

P1 per-field semantic admission is demonstrated at the wire level in
§3.1 Listing 1 (per-field SVAF drift and gate values on the captured
receive-side CMB). The sprint did not surface a clean P1-at-production
event; a schema-contamination catch that appeared at this scale was an
upstream-tooling fault rather than a protocol-level admission decision,
and is therefore excluded from the observations above.

The sprint is evidence of \textbf{cross-session durability}, not of
N-scaling necessity. At N=3 with prompt-level role specialisation, a
single session's attention plus a shared message transport will carry a
bounded task --- we are not claiming the primitives become
\emph{necessary} at N=3. What the primitives deliver is structure that
attention cannot: (i) state that survives session restarts --- multiple
peers in the deployment rejoined fresh context windows without
conversation replay, reconstructing working context from persistent
remix memory rather than from transcripts (O2, P3); (ii) multi-session,
multi-day chains where the total event stream exceeds any single context
window and a rolling attention over a conversation buffer cannot
substitute for evaluated, filtered, provenance-tagged memory; (iii)
receiver-autonomous per-field admission that holds the same semantics
whether the sender is in-session or joined the mesh after a restart
(P1). The boundary where these primitives become necessary rather than
convenient --- the critical N, the critical session-horizon, the
critical attention-vs-persistent-memory trade-off --- is an empirical
question the present deployment does not settle (§5 names it). This is a
reference-deployment report, not a generalisability claim: the
specification is realisable, the four primitives operate jointly as
specified, and cross-session continuity through write-time-filtered
remix memory works in practice --- within the N=3 same-provider scope §5
bounds.

\section{Limitations and Open Work}\label{limitations-and-open-work}

The paper's evidence and specification stop at concrete bounds; we name
those bounds and the open work that extends them. Before listing them we
acknowledge what MMP v0.2.3 \emph{already} specifies, so the remaining
limits are stated honestly rather than inflated.

\textbf{What the MMP specification already addresses.} The spec names
and mitigates cognitive threats directly: state poisoning via
drift-bounded blending, drift manipulation via per-field SVAF, Sybil
attack via aggregation weighted by drift and recency rather than peer
count, and echo loops via lineage parent-check plus organic-mood
isolation. It specifies node identity and authentication (Ed25519),
end-to-end CMB field encryption for relay-traversing frames (X25519 +
XChaCha20-Poly1305, with lineage deliberately in cleartext for DAG
traversal), consent withdrawal as a Layer-2 hard gate that overrides
all coupling, tamper-evident audit via CMB immutability and
content-hash keys, and agent-profile α\_f defaults that implementations
SHOULD expose as configuration. These are specified in MMP v0.2.3 and
are not the paper's contributions to name as limits. What follows are
the genuine bounds.

\textbf{What the empirical deployment does not show.} The §4 operational
deployment is an existence proof at N=3 concurrent agents on a
Mac↔Windows LAN, same-provider (Claude Opus 4.7 family), within a
single-team geography. Three bounds follow directly from that scope.

\begin{itemize}
\tightlist
\item
  \textbf{N-agent convergence.} Pairwise SVAF evaluation scales
  quadratically with peer count; lineage-based echo detection scales
  with ancestor depth (bounded to 50 by §15.2). Whether coordination
  properties observed at N=3 hold at N ≥ 10 remixing meshes is open
  empirical work, and formal convergence analysis of the remix DAG under
  multi-agent churn is an open analytical question.
\item
  \textbf{Cross-provider heterogeneity.} MMP is provider-agnostic by
  construction --- Layer 7 (§14) is implementation-neutral --- but a
  mixed mesh across Claude, GPT, Gemini, and open-source models
  introduces different tokenisation, tool-use conventions, and failure
  modes, and the practical reconciliation is untested.
\item
  \textbf{Comparative evaluation.} §2's gap argument is structural ---
  every surveyed memory substrate is mechanism-differentiated from MMP
  on the retrieval-time-lookup vs acceptance-time-admission axis ---
  not empirical. A head-to-head
  controlled comparison against alternative multi-agent frameworks on
  matched coordination tasks, with benchmarks such as Gaia2 (Froger et
  al.~2026) as natural testbed candidates, would strengthen every claim
  in this paper.
\end{itemize}

\textbf{Adversarial frontiers the spec flags but does not fully close.}
Two items are explicitly marked as future work in the spec itself, and
we surface them here rather than leaving them implicit.

\begin{itemize}
\tightlist
\item
  \textbf{Per-CMB cryptographic signing.} §18.4 states:
  \emph{``Cryptographic CMB signing (future) would make forgery provably
  impossible.''} Without per-CMB signatures, lineage references to
  non-existent keys remain detectable (keys are content hashes), but the
  authenticity of \texttt{createdBy} on a propagated CMB rests on
  transport-level identity rather than a per-CMB signature.
  Forgery-proof provenance awaits the signing extension.
\item
  \textbf{Byzantine collusion under adaptive attack.} The composition of
  drift bounds, per-field admission, and Sybil-insensitive aggregation
  handles common cases, but no formal analysis bounds worst-case
  behaviour under coordinated adversarial peers who are calibrated to
  pass per-field drift thresholds by construction. Formal red-team
  analysis of adaptive adversarial regimes is open.
\end{itemize}

\textbf{What the specification does not formalise.}

\begin{itemize}
\tightlist
\item
  \textbf{Sufficiency and minimality of the primitive set.} §3.5
  conjectures that CAT7 + SVAF + lineage + remix jointly satisfy P1, P2,
  P3 for any assembly qualifying as semantic infrastructure. We do not
  prove this set is unique or minimal. An information-theoretic
  characterisation of the primitives as the minimum decomposition
  distinguishing evaluative cognition from message-passing is open
  analytical work.
\item
  \textbf{Learned or topic-dependent α\_f schedules.} §19.3 specifies
  per-profile static α\_f weights and recommends implementations expose
  them as configuration; §11 specifies Closed-form Continuous-time (CfC;
  Hasani et al.~2022) hidden-state adaptation through feedback
  neuromodulation (fast- and slow-τ neurons). Learning the α\_f weights
  themselves --- RL-driven schedules aligned with Hu et al.~(2025)'s
  RL-driven-memory frontier, or topic-dependent dynamic weighting where
  an auditor shifts α\_f between \texttt{commitment} on scope
  discussions and \texttt{motivation} on risk discussions --- is out of
  scope for the current spec.
\item
  \textbf{Hallucination-robustness beyond source-traceability.} One of
  Hu et al.~(2025)'s three trustworthy-memory pillars (alongside privacy
  preservation and explainability). MMP's lineage DAG provides the
  precondition --- per-claim provenance that supports post-hoc auditing
  --- but does not itself detect or prevent hallucinated content.
  Hallucination-aware per-field admission, or lineage-based
  cross-validation across independent derivations, is open.
\end{itemize}

\textbf{What the integrated layer does not itself provide.} The four
primitives are the substrate on which adjacent capabilities can
accumulate, but the substrate is not the capability.

\begin{itemize}
\tightlist
\item
  \textbf{Experiential-tier playbook induction across the mesh.} MMP
  supplies the substrate for playbooks (reusable strategy traces) to
  accumulate across agents via lineage + remix, rather than privately
  within one. MMP does not perform the \emph{induction} step ---
  extracting a reusable playbook from past CMB chains, in the sense Wang
  et al.~(2024) Agent Workflow Memory does within a single agent ---
  which is downstream work MMP enables.
\item
  \textbf{Composition with existing memory substrates.} MMP supplies
  the inter-agent coordination contract; existing memory stores
  (MemGPT, Mem0, A-MEM, Agent Workflow Memory, Reflexion) can sit
  inside each MMP node as that node's local long-term memory layer.
  MMP does not prescribe the composition patterns --- which substrate
  best fits which MMP-node role, and what cross-contamination or
  double-write issues arise when the local store and the mesh
  remix-store diverge --- which remain open.
\end{itemize}

These are the real bounds. What this paper does establish --- a working
and implemented protocol layer carrying evaluated cognitive state
between cross-session multi-agent LLM peers, specified and running in
production --- holds within them. Each bound names concrete empirical,
analytical, or spec-extension work that extends the layer, not work that
should change before the layer is usable.

\section{Conclusion}\label{conclusion}

Teams of LLM agents now collaborate on tasks that span days, weeks,
and many session restarts. What governs whether that collaboration
compounds into collective intelligence, or decays into drift, is
the layer at which agents share, evaluate, and combine each other's
cognitive state. That layer, \emph{semantic infrastructure}, is the
binding constraint on multi-agent LLM systems, and it has been
missing from the stack.

The Mesh Memory Protocol is that layer. Four composable primitives
realise it. CAT7 fixes the seven-field schema that every exchanged
thought must fit. SVAF evaluates each field against the receiver's
role-indexed anchors, solving P1. Inter-agent lineage, carried as
parents and ancestors of content-hash keys, solves P2. Remix stores
only the receiver's own role-evaluated understanding of each
accepted CMB, never the raw peer signal, solving P3. Together they
turn peer-agent signals into durable, grounded, per-field-evaluated
memory that survives restarts: memory is never copied, only remixed
through each receiver's own role-weighted evaluation, so every
session remains an autonomous peer with its own identity and memory
while collaborating with others across the network.

The growing graph of remixes is where collective intelligence
accrues: no single model holds it, no orchestrator manages it, and
it is the protocol that makes it coherent.

\section*{Disclosures}\label{disclosures}

\begingroup
\setlength{\parindent}{0pt}
\setlength{\parskip}{0.5em}
\sloppy
\emergencystretch=6em
\tolerance=9999
\hbadness=10000

\textbf{Funding.} This work was funded by SYM.BOT Ltd. The author
received no external research grants in support of this paper.

\textbf{Conflicts of interest.} The author is the founder and sole
operator of SYM.BOT Ltd, which develops MMP and its reference
implementations (\texttt{@sym-bot/\allowbreak sym},
\texttt{@sym-bot/\allowbreak mesh-channel},
\texttt{sym-core-\allowbreak swift}, MeloTune iOS). The §4 reference deployment is
the author's one-person startup's daily operations on role-specialised
Claude Code sessions, reported as first-party operational observation.

\textbf{Human subjects.} The reported deployments did not involve human
subjects research. The author is the sole human participant in the §4
deployment and consents to the reporting of operational observations;
no other human data is collected or analysed.

\textbf{AI assistance.} This paper was drafted with assistance from
large language model tools (Claude, Anthropic). The author is
responsible for all scientific content, claims, and citations.

\textbf{Reproducibility.} The MMP v0.2.3 specification
(https://sym.bot/spec/mmp, CC BY 4.0) and the open-source SDKs
(Apache 2.0) permit independent implementation and third-party
deployment. The §4 operational observations are reports of a specific
deployment and are not independently reproducible without an equivalent
mesh; readers seeking reproducibility of the protocol behaviours should
run the open-source stack on their own mesh.

\textbf{Data availability.} The MMP v0.2.3 specification is publicly
available at \url{https://sym.bot/spec/mmp} (CC BY 4.0). SYM Node.js
and SYM Swift reference implementations are open source (Apache 2.0).
The sym-mesh-channel plugin is open source at
\url{github.com/sym-bot/sym-mesh-channel} and installable via
\texttt{/plugin marketplace add sym-bot/\allowbreak sym-mesh-channel};
Anthropic Claude Plugin Directory submission review was completed
2026-04-18, with public-registry propagation pending as of this
writing. The training corpus used in the 14-wave sprint is under
embargo; the pre-registration protocol and aggregate wave-level
statistics are available from the author on request.
\endgroup

\section*{References}\label{references}

\begingroup
\setlength{\parindent}{0pt}
\setlength{\parskip}{0.4em}
\everypar{\hangindent=2em\hangafter=1}
\sloppy
\emergencystretch=3em

Anthropic (2024). Model Context Protocol. modelcontextprotocol.io.

Anthropic (2026a). Claude Channels. code.claude.com/docs/en/channels.

Anthropic (2026b). Claude Code Agent Teams: Orchestrate teams of Claude
Code sessions. code.claude.com/docs/en/agent-teams.

Cemri, M. et al.~(2025). Why Do Multi-Agent LLM Systems Fail?
arXiv:2503.13657.

Chase, H. (2024). LangGraph. github.com/langchain-ai/langgraph.

Chhikara, P. et al.~(2025). Mem0: Building Production-Ready AI Agents
with Scalable Long-Term Memory. arXiv:2504.19413.

Du, P. (2026). Memory for Autonomous LLM Agents: Mechanisms, Evaluation,
and Emerging Frontiers. arXiv:2603.07670.

Du, Y. et al.~(2023). Improving Factuality and Reasoning in Language
Models through Multiagent Debate. ICML 2024. arXiv:2305.14325.

Ehtesham, A. et al.~(2025). A Survey of Agent Interoperability
Protocols: Model Context Protocol (MCP), Agent Communication Protocol
(ACP), Agent-to-Agent Protocol (A2A), and Agent Network Protocol (ANP).
arXiv:2505.02279.

Finin, T. et al.~(1994). KQML as an Agent Communication Language. Proc.
CIKM '94, 456-463.

FIPA (2002). FIPA ACL Message Structure Specification. Foundation for
Intelligent Physical Agents, SC00061G. fipa.org/specs/fipa00061.

Froger, R. et al.~(2026). Gaia2: Benchmarking LLM Agents on Dynamic and
Asynchronous Environments. arXiv:2602.11964.

Google (2025). Announcing the Agent2Agent Protocol (A2A).
developers.googleblog.com/en/a2a-a-new-era-of-agent-interoperability/.

Hasani, R., Lechner, M., Amini, A., Rus, D., \& Grosu, R. (2022).
Closed-form Continuous-time Neural Networks. Nature Machine Intelligence
4, 992--1003. arXiv:2106.13898.

Hohpe, G. \& Woolf, B. (2003). Enterprise Integration Patterns:
Designing, Building, and Deploying Messaging Solutions. Addison-Wesley.

Hong, S. et al.~(2023). MetaGPT: Meta Programming for A Multi-Agent
Collaborative Framework. ICLR 2024 (Oral). arXiv:2308.00352.

Hu, Y. et al.~(2025). Memory in the Age of AI Agents: Forms, Functions
and Dynamics. arXiv:2512.13564.

Labrou, Y., Finin, T., \& Peng, Y. (1999). Agent Communication
Languages: The Current Landscape. IEEE Intelligent Systems 14(2), 45-52.

Lessig, L. (2008). Remix: Making Art and Commerce Thrive in the Hybrid
Economy. Penguin Press.

Lewis, P. et al.~(2020). Retrieval-Augmented Generation for
Knowledge-Intensive NLP Tasks. NeurIPS 2020. arXiv:2005.11401.

Liang, T. et al.~(2023). Encouraging Divergent Thinking in Large
Language Models through Multi-Agent Debate. EMNLP 2024.
arXiv:2305.19118.

Liu, J., Zhao, X., Shang, X., \& Shen, Z. (2026). Dive into Claude Code:
The Design Space of Today's and Future AI Agent Systems.
arXiv:2604.14228 {[}cs.SE{]}.

Packer, C. et al.~(2023). MemGPT: Towards LLMs as Operating Systems.
arXiv:2310.08560.

Park, J. S. et al.~(2023). Generative Agents: Interactive Simulacra of
Human Behavior. UIST 2023. arXiv:2304.03442.

Rezazadeh, A. et al.~(2025). Collaborative Memory: Multi-User Memory
Sharing in LLM Agents with Dynamic Access Control. arXiv:2505.18279.

Riedl, C. (2025). Emergent Coordination in Multi-Agent Language Models.
arXiv:2510.05174.

Shinn, N. et al.~(2023). Reflexion: Language Agents with Verbal
Reinforcement Learning. NeurIPS 2023. arXiv:2303.11366.

Sumers, T. et al.~(2024). Cognitive Architectures for Language Agents.
TMLR. arXiv:2309.02427.

Wang, G. et al.~(2023). Voyager: An Open-Ended Embodied Agent with Large
Language Models. TMLR 2024. arXiv:2305.16291.

Wang, Z. et al.~(2024). Agent Workflow Memory. arXiv:2409.07429.

Wu, Q. et al.~(2023). AutoGen: Enabling Next-Gen LLM Applications via
Multi-Agent Conversation. COLM 2024. arXiv:2308.08155.

Xu, W. et al.~(2025). A-MEM: Agentic Memory for LLM Agents.
arXiv:2502.12110.

Xu, H. (2026a). MeloTune: On-Device Arousal Learning and Peer-to-Peer
Mood Coupling for Proactive Music Curation. arXiv:2604.10815.

Xu, H. (2026b). Symbolic-Vector Attention Fusion for Collective
Intelligence. arXiv:2604.03955.

Yu, Z. et al.~(2026). Multi-Agent Memory from a Computer Architecture
Perspective: Visions and Challenges Ahead. arXiv:2603.10062.

\end{document}